\RequirePackage{ifpdf}
\ifpdf
\else
\PassOptionsToPackage{dvipdfmx}{graphicx}
\PassOptionsToPackage{dvipdfmx}{hyperref}
\fi

\documentclass[12pt,a4paper]{article}
\usepackage{jheppub}
\allowdisplaybreaks[4]
\usepackage{amsmath,bm}
\usepackage{subcaption}
\usepackage{slashed}
\usepackage{mathrsfs}

\usepackage{color}
\usepackage{here}

\newcommand{\pmat}[1]{\begin{pmatrix} #1 \end{pmatrix}}

\title{
Refocusing of Wheeler–DeWitt wave functions at inner horizons
}

 \author[a]{Takeshi Chiba,}
 \author[a]{Keiju Murata,}
 \author[b]{and Daiki Saito}
  \affiliation[a]{Department of Physics, College of Humanities and Sciences, Nihon University, Sakurajosui,
 Tokyo 156-8550, Japan}
 \affiliation[b]{Department of Science Education, Ewha Womans University, Seoul 03760, Korea}

\emailAdd{chibatak@gmail.com}
\emailAdd{murata.keiju@nihon-u.ac.jp}
\emailAdd{dsaito@ewha.ac.kr}

\abstract{
We study quantum gravitational effects inside hyperbolic black holes with both outer and inner horizons by solving the Wheeler–DeWitt (WDW) equation in the minisuperspace approximation. The WDW equation contains a tachyonic region where the effective potential becomes negative. We develop a numerical method that consistently evolves the wave function across this region and obtain stable solutions throughout the entire minisuperspace.
For small values of the parameter $\kappa$, which controls the strength of quantum gravitational effects, the wave packet propagates along the classical trajectory with only mild quantum spreading. As $\kappa$ increases, enhanced quantum effects lead to significant spreading of the wave packet during its propagation. Nevertheless, when the initial state is localized near the outer horizon, the wave packet becomes localized again in the vicinity of the inner horizon. We refer to this recovery of localization as a refocusing phenomenon.
This result suggests that, if the geometry is classical near the outer horizon, it becomes classical again near the inner horizon. Within the minisuperspace approximation, inner-horizon formation is not obstructed by quantum gravitational effects.
}

\begin{document}
\maketitle

\section{Introduction}

Black holes with an inner (Cauchy) horizon, such as Reissner–Nordström and Kerr black holes, have long been known to exhibit potential instabilities associated with the infinite blue-shift of infalling perturbations near the Cauchy horizon~\cite{Simpson:1973ua,Chandrasekhar82}. Early studies of mass inflation and Cauchy horizon instability suggested that even small perturbations can grow without bound and destroy the inner horizon structure~\cite{Poisson:1989zz,Poisson:1990eh,Ori:1991zz}, thereby restoring the predictability of classical general relativity. These issues are deeply connected to Penrose's 
strong cosmic censorship conjecture~\cite{1978tpar.book..217P}, 
which broadly asserts that the maximal Cauchy development of generic initial data should be inextendible, ensuring deterministic evolution in Einstein gravity. In recent years, strong cosmic censorship has been actively revisited, particularly in black holes with a positive cosmological constant, where several works have reported possible violations of the conjecture in near-extremal geometries~\cite{Cardoso:2017soq,Dias:2018etb,Dias:2018ufh} and motivated refined or "rough" formulations of strong cosmic censorship~\cite{Dafermos:2018tha}. Despite this progress, analyses incorporating genuine quantum gravitational effects remain relatively limited. In this work, we investigate how quantum gravitational dynamics affect the formation and stability of Cauchy horizons by solving the Wheeler–DeWitt (WDW) equation in a minisuperspace model of black hole interiors.

The WDW equation~\cite{PhysRev.160.1113} provides a canonical framework for describing quantum gravitational dynamics through a wave function of geometry. Since the WDW equation is a functional differential equation on the full superspace of spatial metrics, solving it in general is notoriously difficult. A common approach is therefore to impose spacetime symmetries and truncate the system to a finite-dimensional minisuperspace model; see Ref.~\cite{Halliwell:1989myn} for a comprehensive review. This minisuperspace quantization has been widely applied not only in quantum cosmology but also to black-hole interiors. In particular, many studies have considered black-hole interiors modeled by Kantowski–Sachs-type geometries~\cite{Kantowski:1966te} and investigated whether quantum gravitational effects can resolve or modify classical singularity formation~\cite{Nambu:1987dh,Nakamura:1993nq,Bouhmadi-Lopez:2019kkt,Perry:2021mch,Brahma:2021xjy,Perry:2021udd,Hartnoll:2022snh,Kan:2022ism,Blacker:2023oan,Piazza:2025uxm,Chiba:2025jhz}.
Recent studies have also investigated WDW wave functions in charged black-hole interiors, including charged AdS black holes and Reissner–Nordström geometries, with particular attention to wave packet dynamics and quantum effects on mass inflation~\cite{Blacker:2023ezy,Chien:2025tzm}. 

Interestingly, we find that quantum gravitational effects do not necessarily destroy the inner horizon. In this paper, we investigate quantum gravity in the interior of black holes with hyperbolic horizons by solving the WDW equation in the minisuperspace approximation. By choosing an appropriate range of parameters, the geometry possesses both outer and inner horizons. Since the spacetime is described purely by gravity and the metric depends only on a single radial coordinate (i.e., it is cohomogeneity-1), this setup provides one of the simplest frameworks for investigating quantum gravitational effects associated with inner horizons.

We impose, as an initial condition at the outer horizon, a wave packet localized around the classical geometry. The subsequent evolution of the wave function is governed by the WDW equation, which takes the form of a wave equation with an effective potential. A key difficulty is that the potential becomes negative in a certain region of minisuperspace, leading to a tachyonic region where the wave function grows exponentially. If one naively solves the WDW equation as a standard initial-value problem, this tachyonic instability overwhelms the solution, and even in the semiclassical limit $\hbar \to 0$, the resulting wave packet fails to reproduce the classical trajectory determined by the Einstein equations.

To overcome this problem, we introduce a modified evolution scheme in which the direction regarded as “time” is changed inside the tachyonic region. This prescription allows us to evolve the wave function stably across the entire minisuperspace while preserving the correct semiclassical behavior. Using this method, we numerically construct globally regular solutions to the WDW equation and investigate the propagation of wave packets inside the black-hole interior.

Our main result is the discovery of a refocusing phenomenon: although the wave packet spreads as it propagates away from the outer horizon, it becomes localized again near the inner horizon. This behavior suggests that, within the minisuperspace approximation, quantum gravitational effects do not necessarily obstruct the formation of the inner horizon. Rather, if the geometry is semiclassical near the outer horizon, it can recover semiclassicality again near the inner horizon. We trace the origin of this phenomenon to the structure of the WDW equation in the tachyonic region, where the equation acquires an approximate reflection symmetry.

The rest of this paper is organized as follows. In Sec.~\ref{HADS}, we review hyperbolic AdS black holes and summarize the parameter range in which both outer and inner horizons are present. In Sec.~\ref{WDWBH}, we derive the WDW equation for the black-hole interior within the minisuperspace approximation and discuss its semiclassical solutions. In Sec.~\ref{SolWDW}, we explain our numerical method for solving the WDW equation in the presence of the tachyonic region and present the resulting wave functions. We then demonstrate the refocusing phenomenon near the inner horizon and discuss its interpretation in terms of semiclassicality and expectation values of geometric observables. Finally, Sec.~\ref{Con} is devoted to conclusions and discussions. Technical details of the numerical computation and asymptotic analysis are presented in the appendices.

\section{AdS black holes with hyperbolic horizons}
\label{HADS}

Here, we consider AdS black holes with hyperbolic horizons in four-dimensional spacetime as one of the simplest black hole solutions possessing an inner horizon. The metric is given by
\begin{equation}
    ds^2 = -f(r)dt^2 + \frac{dr^2}{f(r)} + r^2 dH^2 \ ,
    \label{metric:classical}
\end{equation}
where
\begin{equation}
    f(r) = r^2 - 1 - \frac{\mu}{r} \ .
\end{equation}
The metric on the hyperbolic space is given by
\begin{equation}
    dH^2 = d\theta^2 + \sinh^2 \theta \, d\phi^2 \ .
\end{equation}
We work in units where the AdS radius is set to $\ell = 1$ (i.e., $\Lambda = -3$). This metric is a solution to the Einstein equation $R_{\mu\nu}-g_{\mu\nu}R/2-3g_{\mu\nu}=0$. The parameter $\mu$ is proportional to the energy of the spacetime, which is given by~\cite{Emparan:1999pm,Emparan:1999gf}
\begin{equation}
    E=\frac{V_2 \mu}{8\pi G}\ ,
\end{equation}
where $V_2 = \int dH$ and $G$ is the four-dimensional Newton constant. 
In what follows, we will refer to this solution as a hyperbolic black hole.

\begin{figure}
\begin{center}
\includegraphics[scale=0.5]{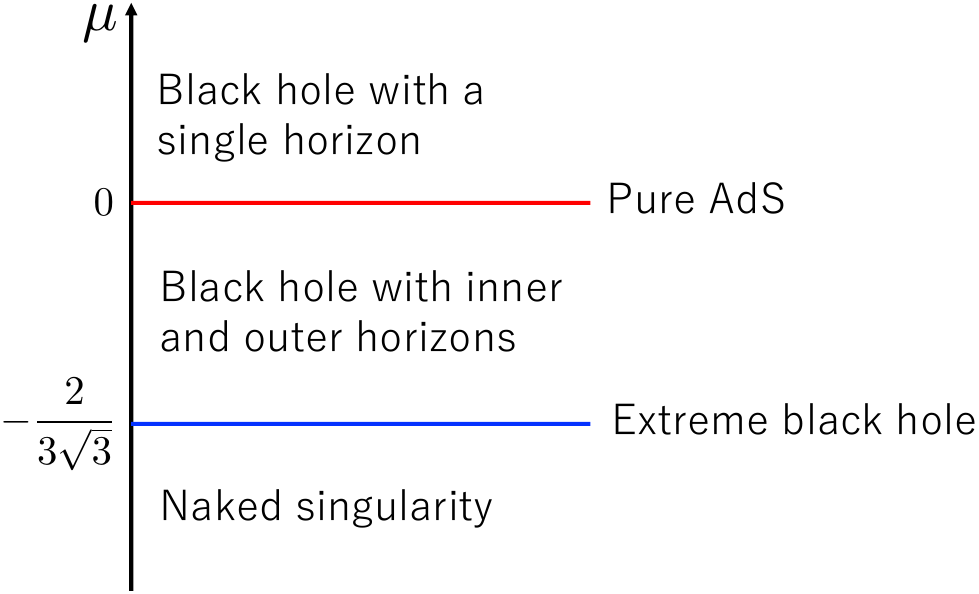}
\end{center}
\caption{Classification of solutions according to the value of $\mu$.
}
\label{muclassify}
\end{figure}

The horizon is determined by the roots of $f(r)$ that satisfy $r > 0$. The condition for the existence of a horizon is given by $\mu \ge -2/(3\sqrt{3})$. For $\mu < -2/(3\sqrt{3})$, the solution describes a naked singularity. In particular, when $\mu = -2/(3\sqrt{3})$, the solution corresponds to an extremal black hole.
For $\mu = 0$, the metric reduces to the hyperbolic patch of pure AdS. For $-2/(3\sqrt{3}) < \mu < 0$, there are two horizons, while for $\mu > 0$, there is a single horizon. Notably, for hyperbolic black holes, there exists a range of negative $\mu$ for which the spacetime remains regular. In this range, the solutions have lower energy than pure AdS. In particular, the extremal solution with $\mu = -2/(3\sqrt{3})$ corresponds to the ground state.
For clarity, the classification of solutions according to the value of the parameter $\mu$ is shown in Fig.~\ref{muclassify}.

In this paper, we focus on the case where both inner and outer horizons are present, namely $-2/(3\sqrt{3}) < \mu < 0$, and study quantum gravitational effects inside the horizons. Let the solutions to $f(r)=0$ be denoted by $r = r_+$ and $r = r_-$ with $r_+ > r_-$. Then, the parameter $\mu$ can be written as
\begin{equation}
    \mu = -r_+(1 - r_+^2) = -r_-(1 - r_-^2) \ .
\end{equation}
In this case, the ranges of the outer and inner horizons are given by $1/\sqrt{3} < r_+ < 1$ and $0 < r_- < 1/\sqrt{3}$, respectively.

\section{Wheeler–DeWitt equation for black hole interiors with inner horizons}
\label{WDWBH}

\subsection{Hamiltonian formulation}

We adopt a metric ansatz \cite{Uglum:1992nc,Gielen:2025ovv,Chiba:2025jhz} that describes both the exterior region of a static black hole and its dynamical interior:
\begin{equation}
    ds^2 = \frac{N(r)^2 U(r)}{V(r)} dr^2 - \frac{V(r)}{U(r)} dt^2 + U(r)^2 dH^2 \ ,
    \label{metric}
\end{equation}
where $U(r)$ and $V(r)$ are functions of the radial coordinate $r$, and $N(r)$ is the lapse function. Substituting this ansatz into the Einstein--Hilbert action with a negative cosmological constant $\Lambda = -3$, supplemented by the Gibbons--Hawking boundary term, the action reduces to an effective one-dimensional form,
\begin{equation}
\begin{split}
    S &= \frac{1}{\kappa_4} \int d^4 x \sqrt{-g} \left( \frac{R}{2} + 3 \right)
    - \frac{1}{\kappa_4} \int d^3 x \sqrt{h}\, K \\
    &= \frac{1}{\kappa} \int dr \left[ \frac{U' V'}{N} + N(-1 + 3 U^2) \right] \ .
\end{split}
\end{equation}
Here, $h$ denotes the determinant of the induced metric on hypersurfaces of constant $r$, and $K$ is the trace of the extrinsic curvature. The constant $\kappa_4 = 8\pi G$ is defined in terms of the four-dimensional Newton constant $G$, while the effective coupling $\kappa$ is given by $\kappa = \kappa_4 / \int dt \, dH$. A prime denotes differentiation with respect to $r$.

The conjugate momenta associated with $U$ and $V$ are obtained as
\begin{equation}
    \pi_U = \frac{1}{\kappa} \frac{V'}{N} \ , 
    \qquad
    \pi_V = \frac{1}{\kappa} \frac{U'}{N} \ .
\end{equation}
From these, the Hamiltonian density takes the form
\begin{equation}
    \mathcal{H} = \pi_U U' + \pi_V V' - \mathcal{L}
    = \kappa N \left[ \pi_U \pi_V - \frac{1}{\kappa^2}(-1 + 3U^2) \right] \ ,
\end{equation}
which leads to the Hamiltonian constraint $\mathcal{H} \approx 0$. The classical dynamics is entirely governed by this constraint.

We introduce a metric in the minisuperspace spanned by $(U,V)$,
\begin{equation}
    g_{AB} = \pmat{0 & -\tfrac{1}{2} \\ -\tfrac{1}{2} & 0} \ , 
    \qquad
    g^{AB} = \pmat{0 & -2 \\ -2 & 0} \ ,
\end{equation}
with indices $A,B = U,V$. The normalization is chosen such that the corresponding line element of the minisuperspace takes the standard two-dimensional Minkowski form,
\begin{equation}
    ds_{\mathrm{MS}}^2 = - dU \, dV \ .
    \label{dsMS}
\end{equation}
Then, the Hamiltonian constraint can be written as
\begin{equation}
    -\frac{1}{4} g^{AB} \pi_A \pi_B - \frac{1}{\kappa^2}(-1 + 3U^2) = 0 \ .
\end{equation}

\subsection{Classical solution}

\begin{figure}
\begin{center}
\includegraphics[scale=0.5]{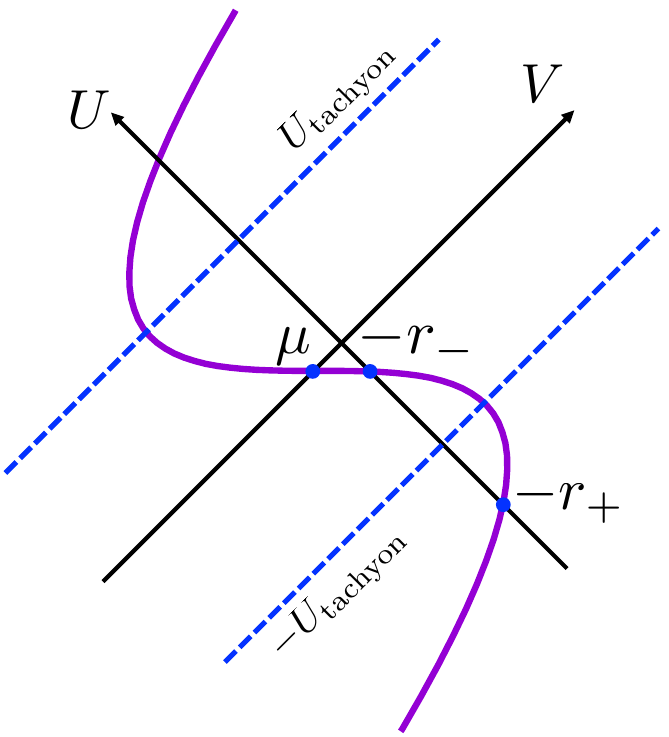}
\end{center}
\caption{Trajectory of the classical solution in minisuperspace. Since both outer and inner horizons are present, the curve intersects the negative $U$-axis twice. Although the classical solution is originally defined only in the region $r > 0$ (i.e., $U < 0$), it is extended into the region $r < 0$ (i.e., $U > 0$). 
}
\label{classicaltraj}
\end{figure}

For the hyperbolic black hole (\ref{metric:classical}), the metric components can be explicitly identified as\footnote{
There is an ambiguity in choosing either $U(r) = r$ or $U(r) = -r$. 
Since $-r$ can be interpreted as a time coordinate in the region $r_- < r < r_+$, we adopt the convention $U = -r$.
}
\begin{equation}
   U(r) = -r \ ,\quad V(r) = -r f(r) \ ,\quad N(r) = 1 \ .
    \label{UVclassical}
\end{equation}
Eliminating the parameter $r$, the classical trajectory in the $(U,V)$ minisuperspace is given by
\begin{equation}\label{eq:classical-trajectory}
V = U^3 - U + \mu \equiv F(U) \, .
\end{equation}
In particular, the locations of the black hole singularity and the inner and outer horizons are mapped to
\begin{equation}
(U,V) = (0,\mu)\quad (r \to 0)\ ,\qquad (U,V) = (-r_\pm ,0)\quad (r \to r_\pm)\,.
\end{equation}
A schematic picture of this classical trajectory is shown in Fig.~\ref{classicaltraj}.
We focus on the case where both outer and inner horizons are present. In this case, the curve intersects the negative $U$-axis twice, at $U = -r_\pm$. The singularity at $r = 0$ corresponds to the $V$-axis, and the curve intersects it at $V = \mu$. Although the classical solution is originally defined only in the region $r > 0$ (i.e., $U < 0$), we extend it and plot it also in the region $r \le 0$ (i.e., $U \ge 0$). As will be seen later, upon solving the WDW equation, the wave function can be naturally extended to the region $U \ge 0$.

We define the turning points of the classical solution, satisfying $F'(U)=0$, as $U = \pm U_\text{tachyon}=\pm 1/\sqrt{3}$. In the region $|U| < U_\text{tachyon}$, the curve becomes spacelike (tachyonic). Special care must be taken in treating this region when solving the WDW equation. This point will be discussed in detail in Sec.~\ref{sec:tachyonic}.

\subsection{Canonical quantization}

We now implement canonical quantization by promoting the phase space variables to operators acting on the wave function $\Psi(U,V)$. The Hamiltonian constraint $\mathcal{H}\,\Psi = 0$ is then promoted to the WDW equation. To fix the operator ordering ambiguity, we adopt a prescription that yields the Laplacian associated with the minisuperspace metric $g_{AB}$ \cite{Hawking:1985bk}. In this approach, the kinetic term is replaced as
\begin{equation}
    g^{AB}\pi_A \pi_B \;\longrightarrow\; -(-g)^{-1/2}\partial_A \left[ (-g)^{1/2} g^{AB} \partial_B \right] \ ,
\end{equation}
where we take the unit of $\hbar = 1$. Then, the WDW equation becomes
\begin{equation}
    \left[  \frac{\partial^2}{\partial U \partial V} + \frac{1}{\kappa^2}\mathcal{V}(U)\right] \Psi(U,V) = 0\ ,\quad \mathcal{V}(U)=-1 +3 U^2
    \label{eq:wdw}
\end{equation}
Here, $\kappa$ plays the role of an expansion parameter in the semiclassical approximation, and the classical limit is obtained in the limit $\kappa \to 0$. When $\hbar$ is restored, the effective expansion parameter is given by $\hbar \kappa \propto \hbar G$, which controls the magnitude of quantum gravitational corrections \cite{Chiba:2025jhz}.

\subsection{Semi-classical solutions}

To extract the semiclassical spacetime dynamics from the WDW equation, we employ the WKB ansatz
\begin{equation}
    \Psi(U,V) = A(U,V) \exp\left( \frac{i}{\kappa} S(U,V) \right)\,,
\end{equation}
where $S(U,V)$ denotes the classical action, and $A(U,V)$ is a slowly varying amplitude. 
Substituting this ansatz into the WDW equation \eqref{eq:wdw} and performing an expansion in powers of $\kappa$, the leading-order contribution ($\mathcal{O}(\kappa^{-1})$) yields the Hamilton--Jacobi equation
\begin{equation}\label{eq:Hamilton-Jacobi}
 \frac{\partial S}{\partial U} \frac{\partial S}{\partial V}
 = -1 + 3 U^2 \,.
\end{equation}
By adopting a separable ansatz of the form $S = S_U(U) + S_V(V)$, the solution to the above equation can be obtained as
\begin{equation}
    S(U,V) = -c \left(U^3-U \right) - \frac{V}{c} + \text{const} \,,
\end{equation}
where $c$ is a separation constant. 
Thus, the semiclassical solution is given by
\begin{equation}
    \Psi \sim \exp\left[ -\frac{i}{\kappa} \left\{
    c \left(U^3-U\right) + \frac{V}{c}
   \right\}\right]\,.
    \label{solution:wkb}
\end{equation}
In fact, (\ref{solution:wkb}) is the exact solution to the WDW equation~(\ref{eq:wdw}) \cite{Uglum:1992nc,Gielen:2025ovv}.  As shown in Ref.~\cite{Chiba:2025jhz}, the constant $c$ corresponds to the freedom of rescaling the coordinate $t$ in the classical solution. In the following, we consider the case $c = 1$, which corresponds to the standard scaling of time.

\section{Solutions to the Wheeler–DeWitt equation}
\label{SolWDW}

In order to investigate how quantum effects become more pronounced as $\kappa$ is increased, we consider a wave packet initially localized around the semiclassical solution and solve the WDW equation directly and numerically to study how the wave packet evolves across the tachyonic region. Although the WKB solution~(\ref{solution:wkb}) obtained in the previous section corresponds to a plane wave and is therefore not suitable for describing such a localized wave packet, its phase factor provides a useful basis for constructing semiclassical initial data.

\subsection{Solving the Wheeler–DeWitt Equation in the presence of a tachyonic region}
\label{sec:tachyonic}

Here, we explain a method to solve the WDW equation~(\ref{eq:wdw}). Although this equation appears at first glance to be a standard wave equation with a potential term, 
one must take into account the fact that the potential $\mathcal{V}$ becomes negative in the region $|U| < U_\text{tachyon} = 1/\sqrt{3}$.  

\begin{figure}
\begin{center}
\includegraphics[scale=0.5]{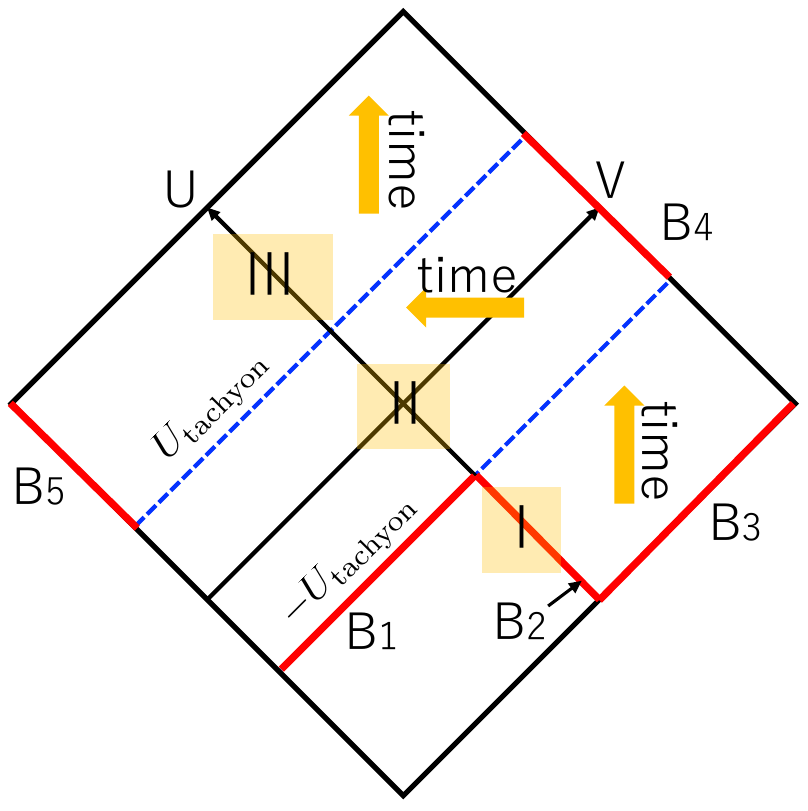}
\end{center}
\caption{Penrose diagram of the minisuperspace. The regions $U \le -U_{\text{tachyon}}$, $|U| < U_{\text{tachyon}}$, and $U \ge U_{\text{tachyon}}$ are referred to as regions I, II, and III, respectively. In regions I and III, the WDW equation is solved by evolving upward, while in region II it is solved by evolving toward the left, each direction being regarded as ``time.'' Initial conditions for solving the WDW equation are specified on the null boundaries $B_1, \cdots, B_5$.
}
\label{regionI-III}
\end{figure}

Fig.~\ref{regionI-III} shows the Penrose diagram of the minisuperspace~(\ref{dsMS}). As illustrated in this figure, we refer to the regions $U \le -U_\text{tachyon}$, $|U| < U_\text{tachyon} $, and $U > U_\text{tachyon} $ as region I, region II, and region III, respectively.  
In region II, the potential is negative, leading to a tachyonic instability, and the wave function grows exponentially. As a result, if one attempts to solve the WDW equation as a standard initial value problem evolving ``from bottom to top'' (i.e., by regarding $T = U + V$ as a time coordinate), the numerical computation becomes unstable and eventually breaks down~\cite{Chiba:2025jhz}. This feature is also reflected in the classical trajectory obtained in Eq.~(\ref{eq:classical-trajectory}). As seen in Fig.~\ref{classicaltraj}, in region II, the trajectory effectively behaves like that of a ``tachyon,'' in the sense that it becomes spacelike, corresponding to motion exceeding the speed of light.

In the presence of a tachyonic instability, the classical trajectory can no longer be recovered from the WDW equation in the tachyonic region, even when the quantum-effect parameter $\kappa$ is taken to be sufficiently small. 
In this paper, we avoid this problem by reversing the direction of ``time'' in region II. Specifically, by performing the change of variables $V = -V'$ in the WDW equation~(\ref{eq:wdw}), we obtain
\begin{equation}
\left[ \frac{\partial^2}{\partial U \, \partial V'} - \frac{1}{\kappa^2}\mathcal{V}(U)\right] \Psi(U,V') = 0 \ ,
\label{wdwprime}
\end{equation}
which flips the sign of the potential term. Therefore, by solving the WDW equation ``from right to left'' (i.e., by regarding $T' = U + V' = U - V$ as a time coordinate), the tachyonic instability can be avoided.

Since we employ a nontrivial procedure in which the direction of time is changed depending on the region, special care must be taken in specifying the initial surfaces. As shown in Fig.~\ref{regionI-III}, we introduce boundary surfaces $B_1, B_2, \cdots, B_5$ and impose initial conditions on each of them.  
For clarity, the boundary surfaces shown in Fig.~\ref{regionI-III} are defined by
\begin{equation}
\begin{split}
B_1 &= \{(U,V)\mid U=-U_{\rm tachyon},\; V<0\},\\
B_2 &= \{(U,V)\mid U\le -U_{\rm tachyon},\; V=0\},\\
B_3 &= \{(U,V)\mid U=-\infty,\; V\geq 0\},\\
B_4 &= \{(U,V)\mid |U|<U_{\rm tachyon},\; V=\infty\},\\
B_5 &= \{(U,V)\mid  U\ge U_{\rm tachyon},\; V=-\infty\}.
\end{split}
\end{equation}

We choose the initial condition on $B_2$ so that it matches the semiclassical solution~(\ref{solution:wkb}) at the outer horizon ($V = 0$). Concretely, we set
\begin{equation}
\Psi\big|_{B_2} = \exp\left(-\frac{(U + r_+)^2}{2\sigma^2}\right)
\exp\left[ -\frac{i}{\kappa} \left( U^3 - U \right)\right] \equiv \Psi_{\textrm{ini}} \,,
\label{init}
\end{equation}
while on the other boundaries $B_1, B_3, B_4, B_5$, we simply impose $\Psi = 0$ for simplicity.

We now examine in detail how the solution is determined in each region. We begin with region I. Under the initial conditions imposed on $B_2$ and $B_3$, solving the WDW equation~(\ref{eq:wdw}) determines the solution in the region with $V \ge 0$. 
Likewise, in the region with $V < 0$, the solution can be determined by solving the WDW equation~(\ref{eq:wdw}) backward under the initial conditions imposed on $B_1$ and $B_2$.
We next turn to region II. From the solution obtained in region I and the initial condition imposed on $B_1$, the wave function at $U = -U_\text{tachyon}$ is already determined. Using this data together with the initial condition on $B_4$, we solve the WDW equation with a reoriented time direction~(\ref{wdwprime}). As a result, the solution in region II can be determined. Finally, we turn to region III. From the solution obtained in region II, the data at $U = U_{\text{tachyon}}$ are already determined, and the initial condition on $B_5$ is also specified. Using these initial data, we solve the WDW equation~(\ref{eq:wdw}), which determines the solution in this region as well. 
Following this procedure, the solution is determined throughout the entire minisuperspace.
In practice, it is not possible to place the boundaries $B_3, B_4, B_5$ at null infinity as depicted in Fig.~\ref{regionI-III}. Instead, they are located at sufficiently large distances. The technical details of this implementation are explained in Appendix~\ref{numerical}. We also describe the discretization scheme used in the numerical computations in this appendix.

In general, the initial conditions for the WDW equation are not uniquely determined \emph{a priori}, and alternative choices are possible. In this work, however, we adopt the above initial condition as a working hypothesis for constructing the quantum theory inside the black hole. 
Our guiding principle is that the wave function should reproduce the classical trajectory in the semiclassical limit $\kappa\to0$. In particular, we impose $\Psi=0$ on $B_1$, since a nontrivial boundary value there would generate a finite wave amplitude away from the classical trajectory even in the semiclassical limit.

\subsection{Numerical solutions of the Wheeler–DeWitt wave function}
\label{numsol}

\begin{figure}[H]
\centering
\begin{subfigure}[b]{0.45\columnwidth}
    \centering
    \includegraphics[width=0.8\columnwidth]{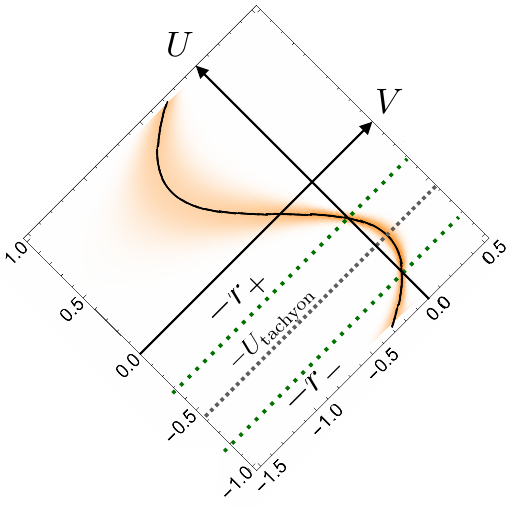}
    \caption{$\kappa=2.5 \times 10^{-3}$}
    \label{NeContoursig002k0001}
\end{subfigure}
\begin{subfigure}[b]{0.5\columnwidth}
    \centering
    \includegraphics[width=0.85\columnwidth]{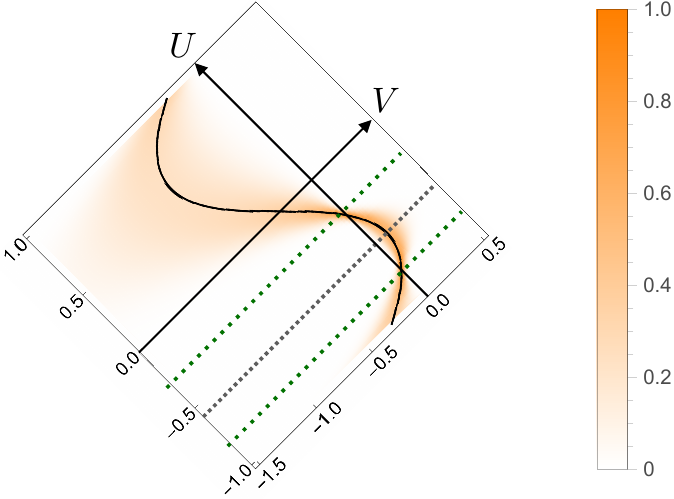}
    \caption{$\kappa=5 \times 10^{-3}$}
    \label{Contour01sig002k0005}
\end{subfigure}
\begin{subfigure}[b]{0.47\columnwidth}
    \centering
    \includegraphics[width=0.8\columnwidth]{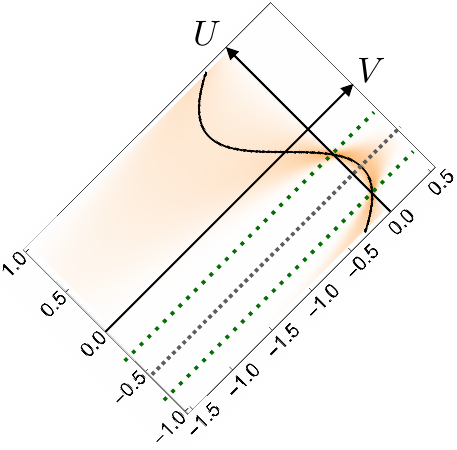}
    \caption{$\kappa=10^{-2}$}
    \label{Contour01sig002k0010}
\end{subfigure}
\begin{subfigure}[b]{0.52\columnwidth}
    \centering
    \includegraphics[width=0.85\columnwidth]{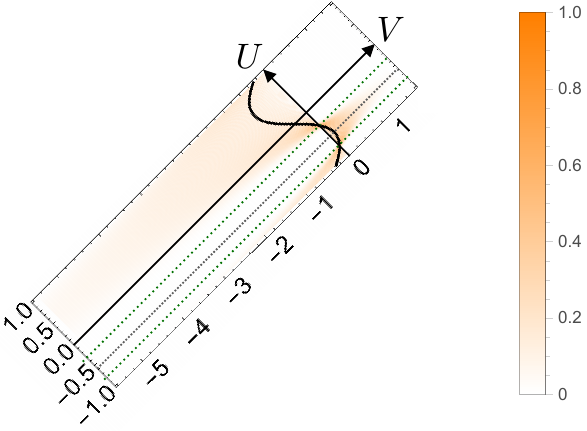}
    \caption{$\kappa=2 \times 10^{-2}$}    \label{Contoursig002k0020}
\end{subfigure}

\caption{Contour of the absolute value of the wave function $|\Psi(U,V)|$. 
We have fixed the parameters as $\sigma=0.02$ and $r_{+}=0.8$, with (a) $\kappa= 2.5\times10^{-3}$, (b) $\kappa=5\times10^{-3}$, and (c) $\kappa=10^{-2}$, and (d) $\kappa=2\times10^{-2}$.
The black curve is the classical trajectory. 
In each panel, $U = -U_\text{tachyon}$ (gray dashed line) and $U = -r_{\pm}$ (green dotted lines) are also shown.}
\label{fig:Cont}
\end{figure}

Figure~\ref{fig:Cont} shows the absolute value of the wave function $|\Psi(U,V)|$ obtained by numerically solving the WDW equation~\eqref{eq:wdw} under the initial condition~\eqref{init} with $\sigma=0.02$, $r_{+} = 0.8$, and $\kappa = 2.5\times10^{-3},\, 5\times10^{-3},\,10^{-2},\,2\times10^{-2}$.
In each panel, we have arranged the boundaries of the computational box so that the wave function decays sufficiently and that the initial conditions for regions II and III are consistent with the assumed values at $B_4$ and $B_5$.  
In this figure, the classical solutions~\eqref{eq:classical-trajectory} are also plotted as the black curves  for reference.

From this figure, the $\kappa$-dependence of the shape of the wave function $|\Psi(U,V)|$ is seen.
For $\kappa=2.5 \times 10^{-3}$, the wave packet sharply traces the classical trajectory  throughout all three regions (I, II, III).
Remarkably, even in the tachyonic region (region II), where the classical solution becomes spacelike, the wave packet remains well-localized and follows the classical path with negligible broadening. 
This demonstrates that our method successfully handles the tachyonic instability and yields a stable evolution across the entire minisuperspace.
As $\kappa$ increases, the quantum gravitational effects become more pronounced. 
For $\kappa=5 \times 10^{-3}$, the wave packet remains centered near the classical trajectory but exhibits 
significant broadening. This spread reflects the increasing uncertainty in the geometry as quantum gravitational effects grow. 
At around $\kappa=10^{-2}$, the wave packet starts to deviate significantly from the classical path for $U>-r_-$, indicating the breakdown of the semiclassical approximation. It is also noteworthy that, although the wave packet originating from the outer horizon ($U=-r_+$) spreads in region I as a consequence of quantum gravitational effects, it refocuses and becomes localized at the inner horizon ($U=-r_-$). This refocusing phenomenon is discussed in detail in Sec. \ref{4.3}.

We also evaluate the behavior of the wave function in the wide range of $\kappa$.
For large $\kappa$, however, a sufficiently large boundary $V_{\text{max}}$ is required to ensure that the wave function has decayed sufficiently, which becomes computationally expensive. 
Therefore, we restrict our analysis of the $\kappa$-dependence to region I, where the computational cost is manageable.

\begin{figure}
\centering
\begin{subfigure}[b]{0.45\columnwidth}
    \centering
    \includegraphics[width=0.75\columnwidth]{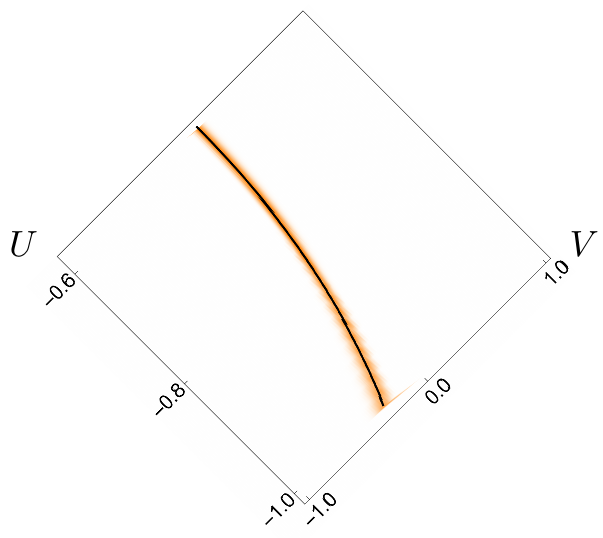}
    \caption{$\kappa=10^{-3}$}
    \label{Contour01sig002k0001}
\end{subfigure}
\begin{subfigure}[b]{0.45\columnwidth}
    \centering
    \includegraphics[width=0.8\columnwidth]{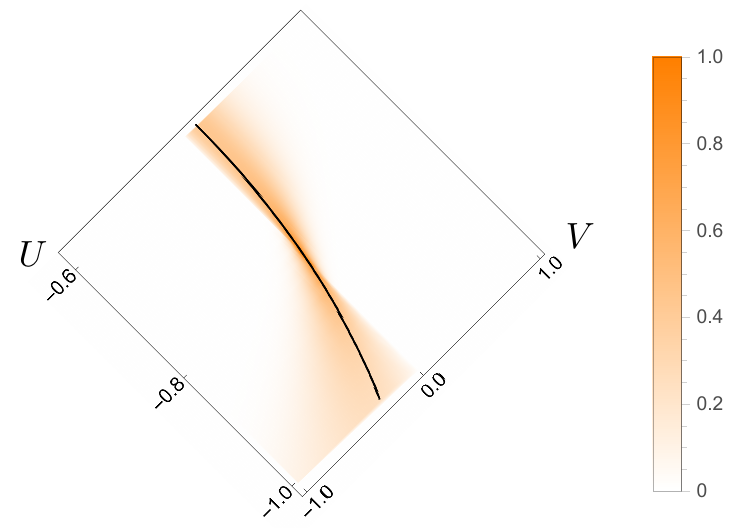}
    \caption{$\kappa=10^{-2}$}
    \label{Contour01sig002k0010}
\end{subfigure}
\begin{subfigure}[b]{0.45\columnwidth}
    \centering
    \includegraphics[width=0.75\columnwidth]{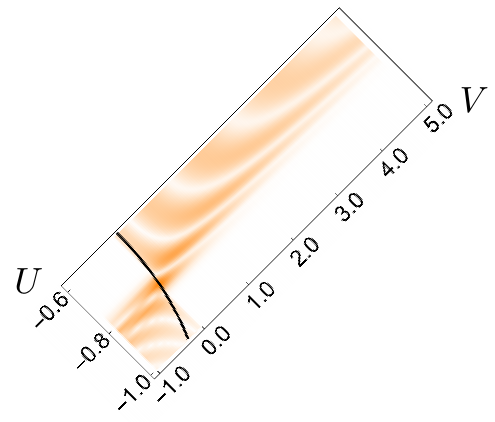}
    \caption{$\kappa=10^{-1}$}
    \label{Contour01sig002k0100}
\end{subfigure}
\begin{subfigure}[b]{0.5\columnwidth}
    \centering
    \includegraphics[width=0.8\columnwidth]{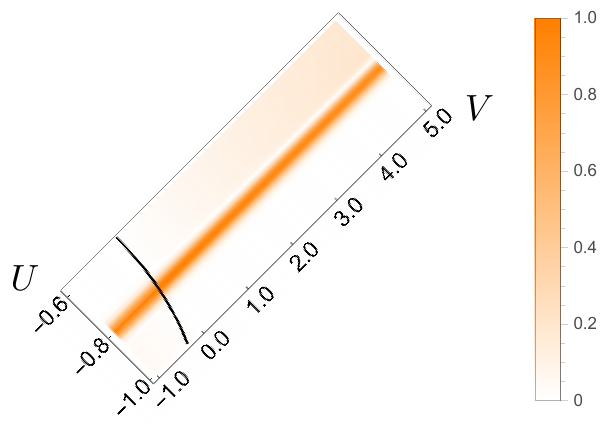}
    \caption{$\kappa=1$}
    \label{Contour01sig002k1000}
\end{subfigure}
\caption{The absolute value of the wave function $|\Psi(U,V)|$ in region I for , with (a) $\kappa= 10^{-3}$, (b) $\kappa=10^{-2}$, and (c) $\kappa=10^{-1}$, and (d) $\kappa=1$.
The boundaries of the computation are fixed as $V_{\text{min}} = -2.0$, $V_{\text{max}} = 5.5$, $U_{\text{min}} = -1.5$, and $U_{\text{max}} = -U_{\text{tachyon}}$, respectively.}
\label{fig:R1}
\end{figure}

Figure~\ref{fig:R1} shows the absolute value of the wave function in region I for $\sigma=0.02$, $r_+=0.8$, and $\kappa = 10^{-3},\,10^{-2},\,10^{-1},$ and $1$.
The qualitative behavior is the same as in Fig.~\ref{fig:Cont}; the peak of the wave function tracks the classical trajectory for $\kappa \ll 1$, and the increasing value of $\kappa$ causes the wave packet to spread. 
For $\kappa=1$, the wave packet no longer follows the classical trajectory. 
Instead, it aligns approximately with the line $U=\mathrm{const}$.
This striking change can be understood from the structure of the WDW equation~\eqref{eq:wdw}.
The potential term is proportional to $\kappa^{-2}$, so that the equation reduces to that of a free wave in the large-$\kappa$ limit.\footnote{
Note, however, that since the wave function decays rapidly for $V\rightarrow\infty$ as long as $\kappa$ is finite, by conservation of the norm (defined in (\ref{inprod_covariant0})
below), it would turn and eventually penetrate into region II (see Appendix \ref{appendixB} for details).}

\subsection{Refocusing of wave function at the inner horizon}
\label{4.3}

In the previous section, we investigated how the value of $\kappa$ affects the dynamics of the wave function. As in Eq.~(\ref{init}), we consider an initial wave packet localized near the outer horizon $(V=0,\; U=-r_+)$. 
Once this physically motivated semiclassical initial condition is specified, the subsequent evolution is uniquely determined by the WDW equation. The refocusing phenomenon therefore emerges naturally during the evolution and is not the result of selecting a particular solution.
As shown in Fig.~\ref{fig:Cont}, the wave packet initially spreads in region I due to finite-$\kappa$ effects. However, as it approaches the inner horizon $(V=0,\; U=-r_-)$, it becomes localized again. We shall refer to this behavior as the \emph{refocusing phenomenon}. In this section, we investigate the mechanism responsible for this refocusing.

This refocusing phenomenon originates from the presence of a tachyonic region in the minisuperspace. Near $U = -U_{\text{tachyon}} = -1/\sqrt{3}$, the potential in the WDW equation can be approximated linearly as
\begin{equation}
\mathcal{V} \simeq -2\sqrt{3}\left(U + U_{\text{tachyon}}\right) \,.
\end{equation}
Under the coordinate transformation that reflects $U$ about
$U=-U_{\text{tachyon}}$,
\begin{equation}
U + U_{\text{tachyon}} = -\left(U' + U_{\text{tachyon}}\right) \,,
\end{equation}
the WDW equation remains invariant. Therefore, in the regime where this linear approximation is valid, the wave function exhibits a symmetric behavior about $U = -U_{\text{tachyon}}$. This approximate symmetry is responsible for the refocusing phenomenon.

Although the wave function of the WDW equation does not admit a standard probabilistic interpretation, we adopt a qualitative viewpoint in which the magnitude of the wave function is regarded as a ``probability-like'' quantity. From this perspective, the spreading of the wave packet in region I due to small-but-finite effects of $\kappa$ indicates that the metric becomes increasingly uncertain under ``time evolution''.
Subsequently, the refocusing observed in region II implies that the quantum state regains a ``classical'' behavior. In other words, if the geometry is classical at the outer horizon, this suggests that it also becomes classical again in the vicinity of the inner horizon.

\subsection{Choice of clock and expectation value of a metric component}

Let us assume that both $\Psi_1$ and $\Psi_2$ satisfy the WDW equation.  
On a hypersurface $\Sigma$, we define the inner product for wave functions as  
\begin{equation}
    (\Psi_1, \Psi_2) = -i \int_\Sigma d\Sigma_A  (\Psi_1^\ast \partial^A \Psi_2-\Psi_2 \partial^A \Psi_1^\ast) \ . \label{inprod_covariant0}
\end{equation}  
Here, we follow the notation of \cite{Poisson:2009pwt} for the surface element $d\Sigma_A$. When the boundary terms can be neglected, the WDW equation implies that the inner product is independent of the choice of hypersurface. 

In our previous work on black holes with a single horizon~\cite{Chiba:2025jhz}, we adopted
$T = U + V$ as the clock. Using this choice, one can naturally define the ``time''
elapsed until singularity formation, and expectation values of geometric observables
evolve in a manner consistent with the classical spacetime picture. In this sense,
the $T$-clock provides a useful interpretation of the quantum dynamics in terms
of semiclassical time evolution.

In the present setup, however, the situation is more subtle due to the existence
of the tachyonic region. As seen in Fig.~\ref{fig:Cont}, the ``center" of the wave packet
generally propagates along a highly curved trajectory in minisuperspace. In particular,
the ``center" of the wave packet can intersect a constant-$T$ hypersurface more than once so that the probability density of 
the wave function can be negative and the probabilistic 
interpretation of the wave function can be problematic \cite{Vilenkin:1988yd}. 
As a consequence, expectation values of observables evaluated on $T=\mathrm{const}$ slices (for example, the expectation value of $X = U - V$) may significantly deviate from the classical prediction even in the semiclassical regime $\kappa \sim 0$. This indicates that the $T$-clock is not well suited
for describing the semiclassical evolution in the presence of the tachyonic region.

For this reason, in this paper we instead adopt $U$ as the clock.
From the viewpoint of the $U$-clock, the ``center" of the wave packet intersects
each constant-$U$ hypersurface only once, and one therefore expects that
expectation values of physical observables reproduce the classical behavior in
the semiclassical limit $\kappa \to 0$.

We consider taking the $U = \text{const}$ surface as the hypersurface $\Sigma$, treating $U$ as the clock. Then, the inner product is given by
\[
(\Psi_1,\Psi_2)=i\int_{-\infty}^\infty dV \left( \Psi_1^\ast \partial_V \Psi_2 - \Psi_2 \partial_V \Psi_1^\ast \right).
\]
However, using this inner product, we generally have
$(\Psi_1, V \Psi_2) \neq (V \Psi_1, \Psi_2)$,
which implies that the operator $V$ is not Hermitian. Consequently, the expectation value of $V$ becomes complex, making its physical interpretation ambiguous.

Here, let us consider the Hermitianized version of the non-Hermitian operator $V$. With respect to the inner product defined above, the Hermitian adjoint of $V$, which satisfies $(\Psi_1,V\Psi_2)=(V^\dagger\Psi_1,\Psi_2)$, is given by
\begin{equation}
V^\dagger = V-(\partial_V)^{-1} \, ,
\end{equation}
where the inverse of the derivative is defined as
\begin{equation}
(\partial_V)^{-1} \Psi(U,V)=\int_{-\infty}^V dV'\,\Psi(U,V') + C(U)\equiv \Phi(U,V)\ .
\end{equation}
Here, $C(U)$ is an integration constant. Although this integration constant is not uniquely fixed, as we will see below, the expectation values of physical observables are independent of its choice. Using the operator $V^\dagger$ introduced above, we obtain
\begin{equation}
    (\Psi_1,V\Psi_2)-(V^\dagger\Psi_1,\Psi_2)
= i\left[((\partial_V)^{-1}\Psi_1^{\ast})\,\Psi_2
\right]_{V=-\infty}^{V=\infty},
\end{equation}
which is a pure boundary term. Since the wave function satisfies $\Psi_2 \to 0$ as $V\to\pm\infty$, this boundary term vanishes.
We then define the Hermitianized operator $V_H$ by
\begin{equation}
V_H = \frac{V+V^\dagger}{2}=V-\frac{1}{2}(\partial_V)^{-1} \, .
\end{equation}
Although this operator is nonlocal, we will ignore this subtlety here and regard $V_H$ as an operator that roughly represents the expectation value of $V$.
The expectation value of this operator is given by
\begin{equation}
\langle V_H \rangle = \frac{\mathrm{Re}(\Psi, V\Psi)}{(\Psi,\Psi)} \, ,
\end{equation}
which is, by construction, real.

Next, to examine the variance of $V_H$, we consider the expectation value of $V_H^2$, which is given by
\begin{equation}
\langle V_H^2 \rangle = \frac{i}{(\Psi,\Psi)}\int_{-\infty}^\infty dV\left[
V^2(\Psi^\ast \partial_V \Psi-\Psi \partial_V \Psi^\ast)
-\frac{1}{4}(\Psi^\ast \Phi-\Psi \Phi^\ast)\right]
\, .
\end{equation}
The above integrand explicitly contains the integral of the wave function, $\Phi$. At first sight, one might therefore suspect that the expectation value depends on the choice of the integration constant $C(U)$. However, this is not actually the case. The terms depending on $C(U)$ take the form
\begin{equation}
    C(U)\int_{-\infty}^{\infty} dV\, \Psi^\ast
    + \text{h.c.}
\end{equation}
Integrating the WDW equation~(\ref{eq:wdw}) over $V$ from $-\infty$ to $\infty$, the $\partial_U\partial_V\Psi$ term becomes boundary terms and vanishes. As a result, one obtains
$\int_{-\infty}^{\infty} dV\, \Psi = 0$. Therefore, all contributions depending on $C(U)$ ultimately drop out. Thus, the expectation value and variance of $V_H$ are well-defined in the sense that they are independent of the ambiguity in the integration constant of the integral operator $(\partial_V)^{-1}$.

\begin{figure}
\centering
\begin{subfigure}[b]{0.45\columnwidth}
    \centering
    \includegraphics[width=0.8\columnwidth]{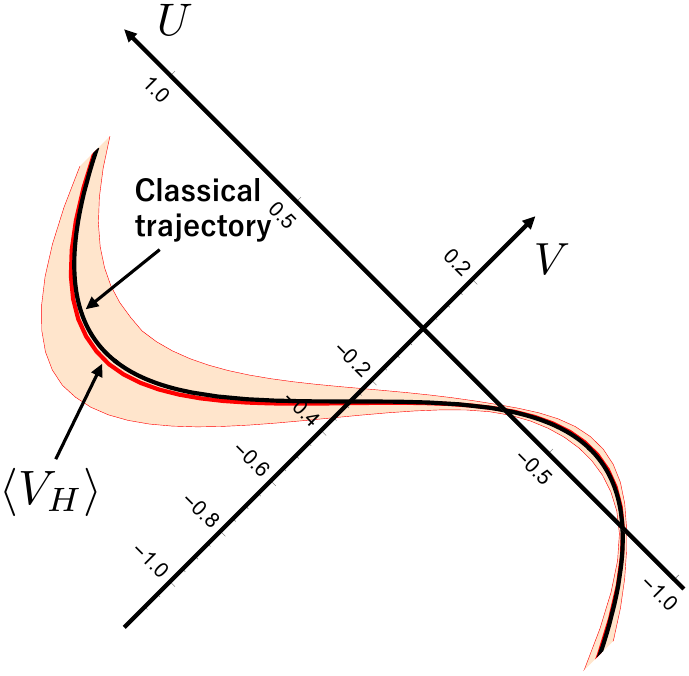}
    \caption{$\kappa=2.5\times 10^{-3}$}
    \label{Vexp_kappa5em3}
\end{subfigure}
\begin{subfigure}[b]{0.45\columnwidth}
    \centering
    \includegraphics[width=0.8\columnwidth]{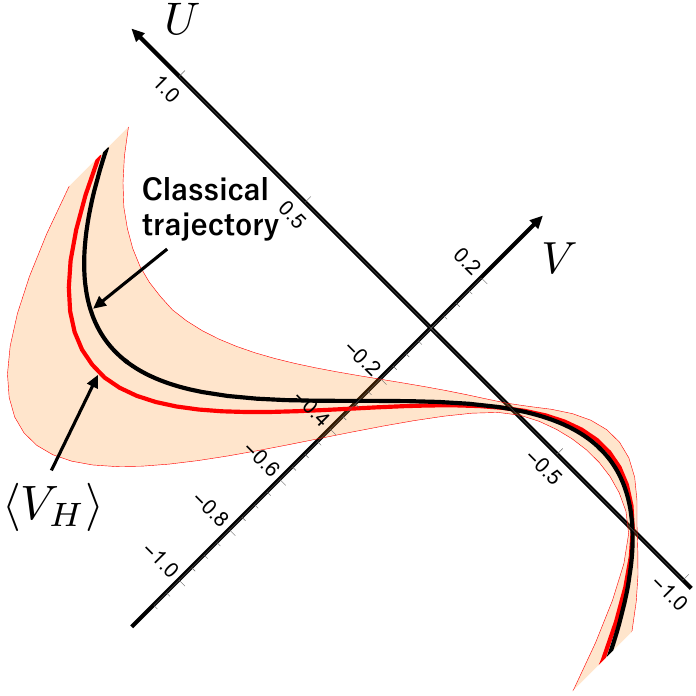}
    \caption{$\kappa=5\times 10^{-3}$}
    \label{Vexp_kappa5em3}
\end{subfigure}
\begin{subfigure}[b]{0.45\columnwidth}
    \centering
    \includegraphics[width=0.8\columnwidth]{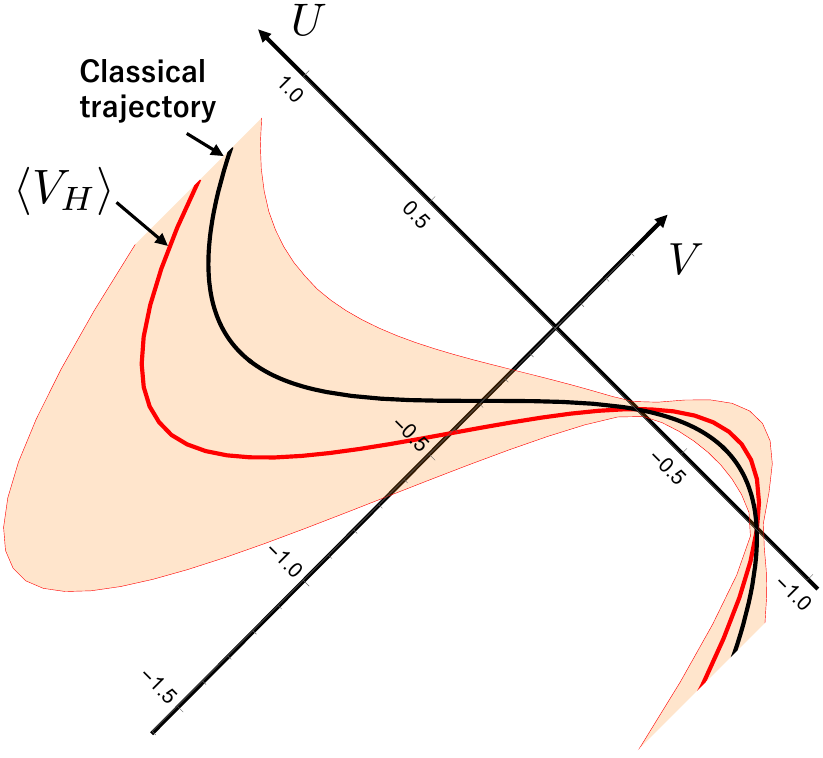}
    \caption{$\kappa=10^{-2}$}
    \label{Vexp_kappa1em2}
\end{subfigure}
\begin{subfigure}[b]{0.45\columnwidth}
    \centering
    \includegraphics[width=0.8\columnwidth]{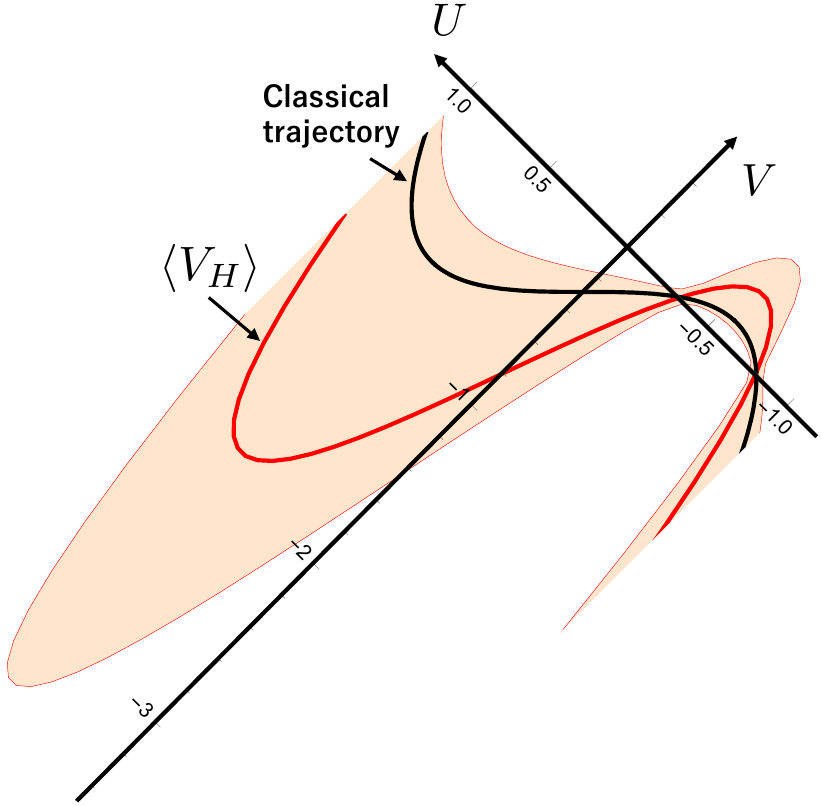}
    \caption{$\kappa=2\times 10^{-2}$}
    \label{Vexp_kappa2em2}
\end{subfigure}
\caption{Expectation value of the Hermitianized operator $V_H$ as a function of $U$ (thick red curve), together with the corresponding classical trajectory (thick black curve). The parameters are $\sigma=0.02$ and $r_+=0.8$, with (a) $\kappa=2.5\times10^{-3}$, (b) $\kappa=5\times10^{-3}$, (c) $\kappa=10^{-2}$, and (d) $\kappa=2\times10^{-2}$. The orange shaded region represents the one-standard-deviation band around the expectation value.}
\label{fig:Vexp}
\end{figure}

We define the variance of the operator $V_H$ as
\begin{equation}
\sigma_{V_H}^2 = \langle V_H^2 \rangle - \langle V_H \rangle^2 \ .
\end{equation}
Figure~\ref{fig:Vexp} shows the expectation value of $V_H$ as a function of $U$ (thick red curve), together with the corresponding classical trajectory (black curve), for $\sigma=0.02$, $r_+=0.8$, and $\kappa = 2.5\times10^{-3},5\times10^{-3},\,10^{-2},\,2\times10^{-2}$. To illustrate the quantum fluctuations of $V_H$, the region satisfying
$|V-\langle V_H\rangle| \le \sigma_{V_H}$
is shaded in orange.

As expected, the expectation value of $V_H$ approaches the classical trajectory as the quantum-effect parameter $\kappa$ decreases.
As $\kappa$ increases, however, the deviation from the classical behavior becomes increasingly significant. Focusing on the variance, one finds that it becomes small near the outer and inner horizons, located at $U=-r_+$ and $U=-r_-$, respectively. The small variance near $U=-r_+$ is naturally understood from the fact that the initial condition is given by a Gaussian wave packet localized around $U=-r_+$. On the other hand, the fact that the variance becomes small again near $U=-r_-$ reflects the refocusing phenomenon discussed in the previous subsection. This is a nontrivial effect originating from the approximate reflection symmetry of the WDW equation. When quantum gravitational effects inside the black hole are taken into account, even if the wave function is initially well localized at the outer horizon, the variance of physical observables grows under the evolution in $U$, indicating that the system becomes increasingly quantum mechanical. However, as the evolution in $U$ proceeds further and approaches the inner horizon at $U=-r_-$, the state becomes classical again in the sense that the variance decreases once more. In this sense, the formation of the outer and inner horizons remains equally deterministic even in the presence of quantum gravitational effects.

\section{Conclusion}
\label{Con}

In this paper, we investigated quantum gravitational effects inside black holes
with both outer and inner horizons by solving the WDW
equation in the minisuperspace approximation. As a concrete setup, we considered
four-dimensional AdS black holes with hyperbolic horizons, which provide one
of the simplest cohomogeneity-1 geometries possessing both outer and inner
horizons. We imposed an initial wave packet localized near the outer horizon
and studied its propagation through the black-hole interior. A central difficulty
in this problem is the existence of a tachyonic region in minisuperspace, where
the effective potential of the WDW equation becomes negative and na\"ive
time evolution becomes unstable. To overcome this issue, we introduced a
modified evolution scheme in which the direction regarded as ``time'' is changed
inside the tachyonic region. Using this prescription, we numerically constructed
globally regular solutions to the WDW equation throughout the entire
minisuperspace. For sufficiently small values of the quantum parameter
$\kappa$, the wave packet follows the classical trajectory with only mild
spreading. More importantly, we found that a wave packet initially localized
near the outer horizon becomes localized again near the inner horizon after
passing through the tachyonic region. We referred to this recovery of
localization as a refocusing phenomenon. This result suggests that, within the
minisuperspace approximation, quantum gravitational effects do not necessarily
obstruct the formation of the inner horizon. Rather, if the geometry is
semiclassical near the outer horizon, it can recover semiclassicality again
near the inner horizon.

There are several important directions for future work. One natural extension is
to investigate other black-hole spacetimes possessing both outer and inner horizons.
A representative example is the Reissner--Nordstr\"om black hole. In that case,
the WDW equation contains not only gravitational degrees of freedom
but also the degree of freedom associated with the Coulomb potential. It would
therefore be interesting to study how the additional gauge-field degree of freedom
modifies the propagation of the wave packet and whether the refocusing phenomenon
persists in such charged black holes.

Another important direction is the study of rotating black holes. In general,
rotating black holes have much lower symmetry, making their quantum theory
extremely difficult to formulate. However, there exist important classes of rotating
black holes that reduce to cohomogeneity-1 systems, such as the BTZ black hole \cite{Banados:1992wn} 
and odd-dimensional rotating black holes with equal angular momenta \cite{Gibbons:2004uw}. In such
cases, it may again become possible to formulate the quantum theory within the
minisuperspace approximation. Through these extensions, it is important to clarify
whether the refocusing phenomenon found in this work is a universal feature of
black holes with inner horizons or merely a special property of the present model.

Another important issue concerns the choice of clock. In this work, we adopt the $U$-clock to define the inner product of the wave function and to evaluate expectation values of the Hermitian operator $V_H$. However, the choice of clock in the WDW equation
is generally ambiguous and remains one of the conceptual subtleties of canonical
quantum gravity. In a series of recent works~\cite{Gielen:2024lpm,Gielen:2025ovv,Gielen:2025grd},
the interior of the AdS Schwarzschild black hole was studied in the context of
unimodular gravity, where the cosmological constant is promoted to a dynamical
variable and a natural time variable conjugate to the cosmological constant emerges.
It would be interesting to apply this framework to black holes with inner
horizons and investigate what kind of quantum evolution is predicted in such a
setup. In particular, it remains an open question whether the refocusing phenomenon
persists when the dynamics is formulated using this alternative notion of time.

\section*{Acknowledgments}
%%%%%%%%%%%%%%%%%%%%%%%%%%%%%%%%%%%%%
We would like to thank Hiroki Matsui for discussions during the early stages of this work. This work is supported in part by Nihon University, and by the National Research Foundation of Korea Grant funded by the Korean Government RS-2024-00336507 (D.S.). 

\appendix

\section{Numerical method}
\label{numerical}

We solve the WDW equation~(\ref{eq:wdw}) numerically using a finite-difference scheme. In this subsection, we describe the numerical method in detail. 
As seen in Eq.~(\ref{solution:wkb}), near the classical limit ($\kappa \to 0$), the phase of the wave function exhibits rapid oscillations. A direct numerical treatment of such behavior requires a mesh size much smaller than the associated wavelength, which significantly increases the computational cost. Thus, we introduce a new function by extracting the rapidly oscillating phase as
\begin{equation}
\Psi(U,V) = \exp\left[ -\frac{i}{\kappa} \left( U^3 - U + V \right) \right] \psi(U,V) \ .
\end{equation}
In terms of $\psi(U,V)$, the WDW equation can be rewritten as
\begin{equation}
\partial_U \partial_V \psi - \frac{i}{\kappa} \left( \partial_U \psi + \mathcal{V}(U)\, \partial_V \psi \right) = 0 \ .
\label{psieq}
\end{equation}

\begin{figure}
\begin{center}
\includegraphics[scale=0.6]{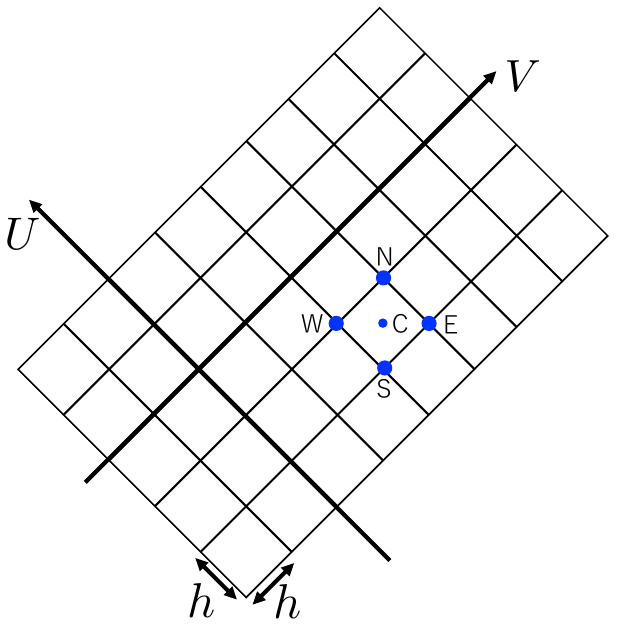}
\end{center}
\caption{Discretization of the minisuperspace for numerical computation.
}
\label{fig:Ndomain}
\end{figure}

As illustrated in Fig.~\ref{fig:Ndomain}, we discretize the $(U,V)$ plane using a uniform grid with spacing $h$. We consider the five lattice points $N$, $E$, $W$, $S$, and $C$ shown in the figure. Finite-difference approximations allow us to express the derivatives at the central point $C$ in terms of the neighboring values at $N$, $E$, $W$, and $S$ as
\begin{equation}
\begin{split}
    &\partial_U\psi\big|_C = \frac{\psi|_N - \psi|_E + \psi|_W - \psi|_S}{2h} \ ,\quad 
    \partial_V\psi\big|_C = \frac{\psi|_N - \psi|_W + \psi|_E - \psi|_S}{2h} \ ,\\
    &\partial_U \partial_V \psi\big|_C = \frac{\psi|_N - \psi|_E - \psi|_W + \psi|_S}{h^2} \ .
\end{split}
\end{equation}
Substituting these expressions into Eq.~(\ref{psieq}) and solving for $\psi|_N$, we arrive at
\begin{equation}
    \psi|_N = \frac{\{1 - \alpha(\mathcal{V} - 1)\}\psi|_W + \{1 + \alpha(\mathcal{V} - 1)\}\psi|_E - \{1 + \alpha(\mathcal{V} + 1)\}\psi|_S}{1 - \alpha(\mathcal{V} + 1)} \ ,
    \label{Ngen}
\end{equation}
where $\alpha = i h/(2\kappa)$.
In region I (see Fig.~\ref{regionI-III}), for $V>0$, the numerical solution can be constructed by successively applying the above update formula using the initial data specified on $B_2$ and $B_3$.

For the lower-right boundary $B_3$, we place it at a location where the Gaussian initial data becomes negligibly small. The numerical solution in this region is obtained in the range $0 < V < V_{\text{max,I}}$. The value of $V_{\text{max,I}}$ is determined by monitoring the wave function at $U = -U_{\text{tachyon}}$ and identifying the point where it becomes sufficiently small. For the region $V<0$ in region I, we use the expression obtained by solving Eq.~(\ref{Ngen}) for $\psi|_S$:
\begin{equation}
\psi|_S=\frac{
    \{1-\alpha(\mathcal{V}-1)\}\psi|_W+\{1+\alpha(\mathcal{V}-1)\}\psi|_E-\{1-\alpha (\mathcal{V}+1)\}\psi|_N}
    {1+\alpha (\mathcal{V}+1)}\ .
\end{equation}
Using this relation together with the wave-function data specified on $B_1$ and $B_2$, the solution in the region $V<0$ can also be determined iteratively.

In region II (the tachyonic region), we solve the WDW equation in the ``right-to-left'' direction in order to avoid the tachyonic instability. Specifically, we use the expression obtained by solving Eq.~(\ref{Ngen}) for $\psi|_W$:
\begin{equation}
\psi|_W
=
\frac{
\{1-\alpha (\mathcal{V}+1)\}\psi|_N
-\{1+\alpha(\mathcal{V}-1)\}\psi|_E+\{1+\alpha (\mathcal{V}+1)\}\psi|_S}{
1-\alpha(\mathcal{V}-1)
}\ .
\end{equation}
Using this relation together with the data specified on $B_1$ and $B_4$, as well as the boundary values obtained from the computation in region I, the solution in region II can be determined.

The solution in region III is determined using Eq.~(\ref{Ngen}), the data specified on $B_5$, and the boundary values of the wave function obtained from the computation in region II.

In practice, the numerical computation is performed in a finite domain 
$U_{\text{min}} < U < U_{\text{max}}$, $V_{\text{min}} < V < V_{\text{max}}$. 
The value of $U_{\text{min}}$ is chosen such that the Gaussian initial data~(\ref{init}) becomes negligibly small. The value of $V_{\text{max}}$ is determined by monitoring the wave function at $U = -U_{\text{tachyon}}$ and identifying the point where it becomes sufficiently small. Similarly, $V_{\text{min}}$ is determined by monitoring the wave function at $U = U_{\text{tachyon}}$ and choosing the point where it becomes sufficiently small.  
The value of $U_{\text{max}}$ can be set depending on how far one wishes to display the solution; in this work, we take $U_{\text{max}} \sim 1$.

\section{Asymptotic behavior of the wave function for $V \to \infty$}
\label{appendixB}

As described in the main text, the initial data in region I are specified on the surface $V=0$. Under the subsequent propagation of the wave, one may ask whether part of the wave escapes to future null infinity $\mathscr{I}^+$ in the minisuperspace, or whether the entire wave eventually passes through $U=-U_{\text{tachyon}}$.
To address this question, in this section we investigate the asymptotic behavior of the wave function in region I in the limit $V\to\infty$. 

To characterize the general form of the solution, let us introduce a Green's function satisfying 
\begin{equation}
    \left[  \frac{\partial^2}{\partial U \partial V} + \frac{1}{\kappa^2}\mathcal{V}(U)\right] G(U,V;U',V') = \delta(U-U')\delta(V-V').
    \label{eq:wdwG}
\end{equation}
The retarded Green function is given as~\cite{Chiba:2025jhz}
\begin{equation}
G(U,V;U',V') = \theta(U-U')\theta(V-V')J_{0}\left(\frac{2}{\kappa}\sqrt{(V-V')(F(U)-F(U'))}\right),
\end{equation}
where $F(U)\equiv \int \mathcal{V}(U) dU = U-U^3$ and 
$J_{n}(x)$ denotes the Bessel function of the first kind.

As shown in Fig.~\ref{regionI-III}, the initial condition in region I is imposed at $V=0$ and $U=U_B$. Here, $U_B$ is chosen to be sufficiently far from the center of the Gaussian so that the initial data~(\ref{init}) becomes negligibly small. In this case, the solution to the WDW equation can be written as
\begin{align}
\Psi(U,V) &= \frac{1}{2}\int_{U_B}^{-U_\text{tachyon}}dU'G(U,V;U',0)\overset{\leftrightarrow}{\partial}_{U'}\Psi(U',0) \nonumber \\
&+ \frac{1}{2}\int_{0}^{\infty}dV'G(U,V;U_B,V')\overset{\leftrightarrow}{\partial}_{V'}\Psi(U_B,V'),
    \label{eq:wdwsol}
\end{align}
where $f\overset{\leftrightarrow}{\partial}_{A}g \equiv f{\partial}_{A}g - g{\partial}_{A}f$.
Since the initial data at $U=U_B$ are set to zero, the second term in the above expression vanishes. Performing integration by parts for the $U'$-integral in the above expression, we obtain
\begin{align}
\Psi(U,V) = \int_{U_B}^{-U_\text{tachyon}}dU'G(U,V;U',0)\partial_{U'}\Psi_{\textrm{ini}}(U') \ .
\end{align}
Since the initial data~(\ref{init}) are sufficiently small at $U=-U_{\text{tachyon}}$ and $U=U_B$, we have neglected the boundary terms.

Let us now examine the behavior of the above expression in the limit $V \to \infty$.
By using the asymptotic form of the Bessel function, $J_{0}(x) \simeq \sqrt{2/(\pi x)}\cos\left(x-\frac{\pi}{4}\right)$ $(x\gg1)$, 
we obtain
\begin{align}
\Psi(U,V) &\simeq \int_{U_B}^{U}dU'\partial_{U'}\Psi_{\textrm{ini}}(U') \nonumber \\
& \times \sqrt{\frac{\kappa}{\pi\sqrt{V(F(U)-F(U'))}}}
\cos\left(\frac{2}{\kappa}\sqrt{V (F(U)-F(U'))}-\frac{\pi}{4} \right). 
\end{align}
By changing the integration variable as $x=(2/\kappa)\sqrt{F(U)-F(U')}$, we obtain
\begin{align}
\Psi(U,V) &\simeq \frac{\kappa^2}{(4\pi^2 V)^{1/4}}\int_0^{X(U)} dx\, \frac{\partial_{U'}\Psi_{\textrm{ini}}}{\mathcal{V}(U')}\cos\left(\sqrt{V}x-\frac{\pi}{4}\right)\ ,
\end{align}
where $X(U)\equiv (2/\kappa) \sqrt{F(U)-F(U_B)}$.
The above expression can be regarded as the Fourier transform of a smooth function multiplied by a rectangular window function. It is known that, for a smooth function $f(x)$, $\int_a^b dx\, f(x)\, e^{ikx} \sim k^{-1}$ ($k \to \infty$). Therefore, in the limit $V \to \infty$, the above expression behaves as
 \begin{equation}
  \Psi(U,V) \sim V^{-3/4} \ .
  \label{Psiatinf}
  \end{equation}

  Here, we define the null hypersurfaces in the minisuperspace by
\begin{equation}
\begin{split}
&\Sigma_\text{in} = \{(U,V)\,|\, U = -U_{\text{tachyon}},\; V > 0\}\ ,\\
&\Sigma_\text{out} = \{(U,V)\,|\, U < -U_{\text{tachyon}},\; V = \infty\}\ ,
\end{split}
\end{equation}
and denote the norm on a hypersurface $\Sigma$ as
\begin{equation}
I_\Sigma=(\Psi,\Psi)|_\Sigma \ .
\end{equation}
By conservation of the norm, one obtains
$I_{B_2}=I_{\Sigma_\text{in}}+I_{\Sigma_\text{out}}$.
On the other hand, as shown in Eq.~(\ref{Psiatinf}), the wave function decays sufficiently rapidly in the limit $V\to\infty$, which implies $I_{\Sigma_\text{out}}=0$.
Therefore, it follows that
\begin{equation}
I_{B_2}=I_{\Sigma_\text{in}}\,.
\end{equation}
This means that, in the sense of the conserved norm, the entire wave eventually passes through $\Sigma_\text{in}$.

When $\kappa$ is large, the wave packet initially appears to propagate almost straight along lines of constant $U$ as shown in Fig.~\ref{fig:R1}. However, as long as $\kappa$ remains finite, the wave ultimately passes through $\Sigma_\text{in}$ after sufficiently long evolution. The exceptional case is the limit $\kappa=\infty$, in which $I_{B_2}=I_{\Sigma_\text{out}}$ 
holds, meaning that the wave reaches $\Sigma_\text{out}$ without passing through $\Sigma_\text{in}$.

%%%%%%%%%%%%%%%%%%%%%%%%%%%%%%%%%%
\bibliography{refs}
%%%%%%%%%%%%%%%%%%%%%%%%%%%%%%%%%%

\end{document}